\documentclass[aps,pra,singlecolumn,superscriptaddress]{revtex4-2}

\usepackage[dvips]{graphicx}
\usepackage[usenames,svgnames]{xcolor}
\usepackage{color}
\usepackage{url}
\usepackage{natbib}
\usepackage[colorlinks=true,linkcolor=magenta,citecolor=magenta,breaklinks=true]{hyperref}

\usepackage{amsmath,amssymb,bm,amsthm,mathtools}

\begin{document}

%\preprint{}

\title{Benchmarking adiabatic transformation by alternating unitaries}


\author{Takuya Hatomura}
\email[]{takuya.hatomura@ntt.com}
%\thanks{}
\affiliation{NTT Basic Research Laboratories \& NTT Research Center for Theoretical Quantum Information, NTT Corporation, Kanagawa 243-0198, Japan}


\date{\today}

\begin{abstract}
Adiabatic transformation can be approximated as alternating unitary operators of a Hamiltonian and its parameter derivative as proposed in a gate-based approach to counterdiabatic driving (van Vreumingen, arXiv:2406.08064). 
In this paper, we conduct numerical benchmarking of this alternating unitary method in a finite-parameter range against adiabatic driving in nonadiabatic timescale. 
We find that the alternating unitary method results in broader distribution on energy eigenstates than that obtained by adiabatic driving, but it has ability to sample low-energy eigenstates when an energy gap of a given Hamiltonian is small. 
It indicates that the alternating unitary method may be able to find good approximate solutions in quantum annealing applied to hard instances. 
\end{abstract}

%\keywords{}

\maketitle


%
%=========================================================================
%
\section{Introduction}

Adiabatic driving is a basic way of quantum control. 
It enables us to track an eigenstate of a Hamiltonian by slowly changing its parameters against an energy gap~\cite{Born1928,Kato1950}. 
Such a control means can be applied to various quantum information processing, e.g., optimization~\cite{Kadowaki1998}, computation~\cite{Farhi2000}, search~\cite{Roland2002}, etc. 
Remarkably, adiabatic quantum computation has the same computational power with standard quantum computation~\cite{Aharonov2007}. 
Moreover, there is numerical evidence that proliferation of errors is suppressed in quantum adiabatic algorithms as opposed to standard quantum algorithms~\cite{Schiffer2024}.

Quantum adiabatic algorithms should potentially be useful, but there are obstacles to exact implementation of them. 
A main problem is the presence of thermal baths in realistic quantum devices. 
A long operation time is required to maintain adiabaticity according to the adiabatic condition~\cite{Messiah1999,Amin2009,Jansen2007,vanVreumingen2023}. 
As a result, quantum adiabatic algorithms suffer from dissipation for a long time and fail in finding exact solutions of given problems. 
However, it should be noted that imperfect implementation of quantum adiabatic algorithms is still fascinating because approximate solutions of some problems can practically be useful~\cite{Lucas2014}.

Introduction of additional control fields is a way of removing the requirement of the adiabatic condition~\cite{Unanyan1997,Emmanouilidou2000}. 
By counteracting diabatic transitions, we can realize fast adiabatic passage. 
Nowadays, such assisted adiabatic control is known as counterdiabatic driving~\cite{Demirplak2003,Demirplak2008,Berry2009}, or more generally, shortcuts to adiabaticity~\cite{Chen2010,Torrontegui2013,Guery-Odelin2019,Hatomura2024}. 
In counterdiabatic driving applied to general quantum many-body systems, time-dependent control of non-local and many-body interactions is required, and thus we usually adopt some approximations~\cite{Sels2017,Claeys2019,Hatomura2021,Takahashi2024,Bhattacharjee2023}. 
Benchmarking tests of approximate counterdiabatic driving have revealed that it has ability to suppress some nonadiabatic transitions in ultra-fast control processes (see Refs.~\cite{Torrontegui2013,Guery-Odelin2019,Hatomura2024} and references therein).

Digitization of adiabatic driving, in which a time-evolution operator is discretized and each time slice is divided into gate operations or implementable components, was discussed as a means of approximate implementation~\cite{Steffen2003,Shapiro2007}, whereas there is room for discussion whether it preserves advantages of adiabatic control or not. 
Recently, it was pointed out that the Trotter error in digitized adiabatic driving shows a scaling advantage against general Hamiltonian simulation~\cite{Kovalsky2023}. 
The idea of digitization was also applied to counterdiabatic driving~\cite{Hegade2021,Hegade2022}. 
Similar to adiabatic driving, a scaling advantage of the Trotter error in digitized counterdiabatic driving was also shown against general Hamiltonian simulation~\cite{Hatomura2022,Hatomura2023}.

More recently, a gate-based algorithm of counterdiabatic driving was proposed~\cite{vanVreumingen2024}. 
In this algorithm, adiabatic transformation is approximated as alternating unitary operators of a given Hamiltonian and its parameter derivative. 
Remarkably, this algorithm has guarantees of success in contrast to famous approximate counterdiabatic driving~\cite{Sels2017,Claeys2019,Hatomura2021,Takahashi2024,Bhattacharjee2023} and digitized counterdiabatic driving with approximations~\cite{Hegade2021,Hegade2022}. 
It was pointed out that its gate complexity evaluated by inequalities is nearly same with digitized adiabatic driving, and thus it cast doubt on advantages of digitized counterdiabatic driving. 
Note that the analysis in Ref.~\cite{vanVreumingen2024} is based on inequalities and they do not discuss actual performance in concrete models.

The purpose of the present paper is to conduct numerical benchmarking of the gate-based counterdiabatic driving, which we call the alternating unitary method, in a finite-parameter range against adiabatic driving in nonadiabatic timescale. 
First, we introduce concise expression of alternating unitary operators which approximates adiabatic transformation. 
Then, we numerically study performance of this alternating unitary method in a finite-parameter range. 
By applying it to a two-level system, we find possibility that the present method gives better results than adiabatic driving when an energy gap is small. 
We reveal that it can actually sample low-energy eigenstates in quantum annealing of the $p$-spin model with $p=3$, which shows the first-order transition with exponential gap closing, while it results in a bad result for $p=2$, with which the $p$-spin model shows the second order transition with polynomial gap closing (per single spin). 
This result indicates that the alternating unitary method may be useful in quantum annealing applied to hard instances.

This paper is constructed as follows. 
In Sec.~\ref{Sec.setup}, we briefly summarize our problem setting and a background. 
In particular, adiabatic gauge potential, which is a generator of energy eigenstates in parameter space, and its regularized version are introduced. 
We introduce the alternating unitary method which approximates adiabatic transformation in Sec.~\ref{Sec.method}. 
Numerical benchmarking of the alternating unitary method is conducted in Sec.~\ref{Sec.num}. 
We summarize the present results in Sec.~\ref{Sec.sum}.

%
%========================================================================
%
\section{\label{Sec.setup}Setup and background}

We introduce a parameterized Hamiltonian $\hat{H}(\bm{\lambda})$ with a vector of parameters $\bm{\lambda}$, and consider state transfer from $|n(\bm{\lambda}_i)\rangle$ to $|n(\bm{\lambda}_f)\rangle$ except for a phase factor, where $|n(\bm{\lambda})\rangle$ is an energy eigenstate of the Hamiltonian. 
The vectors $\bm{\lambda}_i$ and $\bm{\lambda}_f$ represent that of the initial parameters and that of the final parameters, respectively.

One of the ways to realize the above state transfer is adiabatic driving~\cite{Born1928,Kato1950}. 
We express the parameters $\bm{\lambda}$ as $\bm{\lambda}=\bm{\lambda}(s)$, where $s=t/T\in[0,1]$ is the normalized time, $t\in[0,T]$ is the physical time, and $T$ is the operation time. 
The initial and final parameters are denoted by $\bm{\lambda}_i=\bm{\lambda}(0)$ and $\bm{\lambda}_f=\bm{\lambda}(1)$. 
Time evolution under the Hamiltonian $\hat{H}(\bm{\lambda})$ is expressed as $|\psi_n(t)\rangle=\mathcal{T}\exp[-(i/\hbar)\int_0^tdt^\prime\hat{H}\bm{(}\bm{\lambda}(t^\prime/T)\bm{)}]|n(\bm{\lambda}_i)\rangle$. 
A rigorous analysis gives a guarantee of adiabaticity, $|\langle n(\bm{\lambda}_f)|\psi_n(T)\rangle|\ge1-\epsilon$, for the operation time $T\ge T_\mathrm{ad}=\mathcal{O}(\Delta_\mathrm{min}^{-3}\epsilon^{-1})$, where $\Delta_\mathrm{min}$ is the minimum energy gap between the level $E_n(\bm{\lambda})$ and the others~\cite{Jansen2007,vanVreumingen2023}.

Another way of the state transfer from $|n(\bm{\lambda}_i)\rangle$ to $|n(\bm{\lambda}_f)\rangle$ is use of parallel transport by adiabatic gauge potential~\cite{Kolodrubetz2017}
\begin{equation}
|n(\bm{\lambda}+\delta\bm{\lambda})\rangle=\exp\left(-\frac{i}{\hbar}\delta\bm{\lambda}\cdot\hat{\mathcal{A}}(\bm{\lambda})\right)|n(\bm{\lambda})\rangle, 
\label{Eq.para.trans}
\end{equation}
where $\delta\bm{\lambda}$ is infinitesimal displacement and $\hat{\mathcal{A}}(\bm{\lambda})$ is the adiabatic gauge potential
\begin{equation}
\hat{\mathcal{A}}(\bm{\lambda})=i\hbar\sum_{\substack{m,n \\ (m\neq n)}}|m(\bm{\lambda})\rangle\langle m(\bm{\lambda})|\bm{\nabla}n(\bm{\lambda})\rangle\langle n(\bm{\lambda})|. 
\label{Eq.AGP}
\end{equation}
Here, we fix the gauge so that $\langle n(\bm{\lambda})|\bm{\nabla}n(\bm{\lambda})\rangle=0$. 
Note that Eq.~(\ref{Eq.para.trans}) can be regarded as the quench limit of counterdiabatic driving~\cite{vanVreumingen2024}. 
The adiabatic gauge potential (\ref{Eq.AGP}) can be approximated by using the regularized adiabatic gauge potential~\cite{Pandey2020,Bhattacharjee2023}
\begin{equation}
\hat{\mathcal{A}}_\eta(\bm{\lambda})=-\frac{1}{2}\int_{-\infty}^\infty ds\ \mathrm{sgn}(s)e^{-\eta|s|}e^{\frac{i}{\hbar}\hat{H}(\bm{\lambda})s}\bm{(}\bm{\nabla}\hat{H}(\bm{\lambda})\bm{)}e^{-\frac{i}{\hbar}\hat{H}(\bm{\lambda})s}, 
\label{Eq.reg.AGP}
\end{equation}
for small $\eta$ (see, Appendix~\ref{Sec.Appendix.reg.AGP}).

%
%===================================================================
%
\section{\label{Sec.method}Method}

In this section, we introduce the alternating unitary method, which was proposed in Ref.~\cite{vanVreumingen2024}. 
Note that we give more concise expression than the original proposal~\cite{vanVreumingen2024}. 
In this method, we approximate parallel transport (\ref{Eq.para.trans}) as
\begin{equation}
|n(\bm{\lambda}+\delta\bm{\lambda})\rangle\approx\hat{\mathcal{U}}_\eta^M(\bm{\lambda},\delta\bm{\lambda})|n(\bm{\lambda})\rangle, 
\end{equation}
with a unitary operator $\hat{\mathcal{U}}_\eta^M(\bm{\lambda},\delta\bm{\lambda})$ consisting of alternating unitaries of $\hat{H}(\bm{\lambda})$ and $\bm{(}\delta\bm{\lambda}\cdot\bm{\nabla}\hat{H}(\bm{\lambda})\bm{)}$. 
We explain explicit expression and its derivation below.

First, we discretize the regularized adiabatic gauge potential (\ref{Eq.reg.AGP}). 
By applying the trapezoidal rule of integral to the regularized adiabatic gauge potential (\ref{Eq.reg.AGP}), we obtain approximate adiabatic gauge potential
\begin{equation}
\hat{\mathcal{A}}_\eta^M(\bm{\lambda})
=\frac{1}{2\eta M}\sum_{\substack{m=-M+1 \\ (m\neq0)}}^{M-1}\mathrm{sgn}(m)e^{-\frac{i}{\hbar\eta}\mathrm{sgn}(m)\log|m/M|\hat{H}(\bm{\lambda})}\bm{(}\bm{\nabla}\hat{H}(\bm{\lambda})\bm{)}e^{\frac{i}{\hbar\eta}\mathrm{sgn}(m)\log|m/M|\hat{H}(\bm{\lambda})}, 
\label{Eq.app.AGP}
\end{equation}
where $M$ is the number of steps for the trapezoidal rule of integral (see, Appendix~\ref{Sec.disc.reg.AGP} for detailed calculation). 
In the derivation, we avoid introduction of a cutoff by using variable conversion and adopt the trapezoidal rule for simplicity, whereas a cutoff of the integral in Eq.~(\ref{Eq.reg.AGP}) was introduced and weighted sum was considered to discretize the regularized adiabatic gauge potential in the original proposal~\cite{vanVreumingen2024}.

Next, we approximate parallel transport of the energy eigenstate (\ref{Eq.para.trans}) by using the approximate adiabatic gauge potential (\ref{Eq.app.AGP}). 
By considering Trotterization, the unitary operator of parallel transport is approximated as (see, Appendix~\ref{Sec.disc.reg.AGP} for detailed calculation)
\begin{equation}
\hat{\mathcal{U}}_\eta^M(\bm{\lambda},\delta\bm{\lambda})=\hat{u}_\eta^M(\bm{\lambda})\hat{U}_\eta^M(\bm{\lambda},\delta\bm{\lambda})\hat{u}_\eta^M(\bm{\lambda}),
\label{Eq.approx.parallel}
\end{equation}
where 
\begin{equation}
\hat{U}_\eta^M(\bm{\lambda},\delta\bm{\lambda})=\left(\prod_{m=-1}^{-M+1}e^{-\frac{i}{2\hbar\eta M}\bm{(}\delta\bm{\lambda}\cdot\bm{\nabla}\hat{H}(\bm{\lambda})\bm{)}}e^{-\frac{i}{\hbar\eta}\log\bm{(}m/(m-1)\bm{)}\hat{H}(\bm{\lambda})}\right)\left(\prod_{m=M-1}^1e^{-\frac{i}{\hbar\eta}\log\bm{(}m/(m+1)\bm{)}\hat{H}(\bm{\lambda})}e^{\frac{i}{2\hbar\eta M}\bm{(}\delta\bm{\lambda}\cdot\bm{\nabla}\hat{H}(\bm{\lambda})\bm{)}}\right),
\label{Eq.unitary1}
\end{equation}
and
\begin{equation}
\hat{u}_\eta^M(\bm{\lambda})=e^{\frac{i}{\hbar\eta}\log(1/M)\hat{H}(\bm{\lambda})}. 
\label{Eq.unitary2}
\end{equation}
That is, the unitary operator $\hat{\mathcal{U}}_\eta^M(\bm{\lambda},\delta\bm{\lambda})$ is alternating unitaries of $\hat{H}(\bm{\lambda})$ and $\bm{(}\delta\bm{\lambda}\cdot\bm{\nabla}\hat{H}(\bm{\lambda})\bm{)}$.

Now we discuss an effective operation time of the unitary (\ref{Eq.approx.parallel}). 
We regard the coefficients of $(i/\hbar)\hat{H}(\bm{\lambda})$ and $(i/\hbar)\bm{(}\delta\bm{\lambda}\cdot\bm{\nabla}\hat{H}(\bm{\lambda})\bm{)}$ in the unitaries (\ref{Eq.unitary1}) and (\ref{Eq.unitary2}) as an effective operation time. 
There may be room for discussion in the latter definition, but it is not a dominant term. 
The total effective operation time is given by
\begin{equation}
T_\mathrm{eff}=\frac{4L}{\eta}\log M+\frac{L}{\eta}\left(1-\frac{1}{M}\right), 
\end{equation}
where $L$ is the number of parameter steps, i.e., $L=|\mathcal{C}|/|\delta\bm{\lambda}|$ for a path $\mathcal{C}$ from $\bm{\lambda}_i$ to $\bm{\lambda}_f$. 
Here, $|\mathcal{C}|$ represents the length of the path $\mathcal{C}$. 
That is, the effective operation time is proportional to the number of parameter steps $L$, inversely proportional to the regularizer $\eta$, and logarithmically proportional to the integer $M$. 
Note that we can further reduce this total effective operation time by combining different order of decomposition and introducing additional approximation (see Appendix~\ref{Sec.reduc.time}).

%
%========================================================================
%
\section{\label{Sec.num}Numerical simulation}

Now we numerically study performance of the alternating unitary method. 
In this section, we set $\hbar=1$.

%
%----------------------------------------------------------------------
%
\subsection{Two-level system}

First, we consider a spin flip of a two-level system described by a Hamiltonian
\begin{equation}
\hat{H}(\bm{\lambda})=-h^x\hat{X}-h^z\hat{Z}, 
\end{equation}
where $\hat{X}$ and $\hat{Z}$ are the Pauli-X operator and the Pauli-Z operator, respectively. 
Parameters are a transverse field and a longitudinal field, $\bm{\lambda}=(h^x,h^z)$, and we can realize a spin flip by $\bm{\lambda}_i=(0,h_0^z)$ and $\bm{\lambda}_f=(0,-h_0^z)$ with a certain constant $h_0^z$. 
We assume the following parameter schedule
\begin{equation}
h^x(s)=h_0^x\sin(\pi s),\quad h^z(s)=h_0^z\cos(\pi s), 
\end{equation}
for state transfer by adiabatic driving. 
Here, $h_0^x$ is also a constant and we consider cases with $h_0^x/h_0^z\le1$ ($h_0^x>0$ and $h_0^z>0$). 
By varying $h_0^x$, we can change the size of the energy gap $\Delta_\mathrm{min}=2h_0^x$. 
We also adopt the same parameter path in the alternating unitary method.

\begin{figure}
\includegraphics[width=7cm]{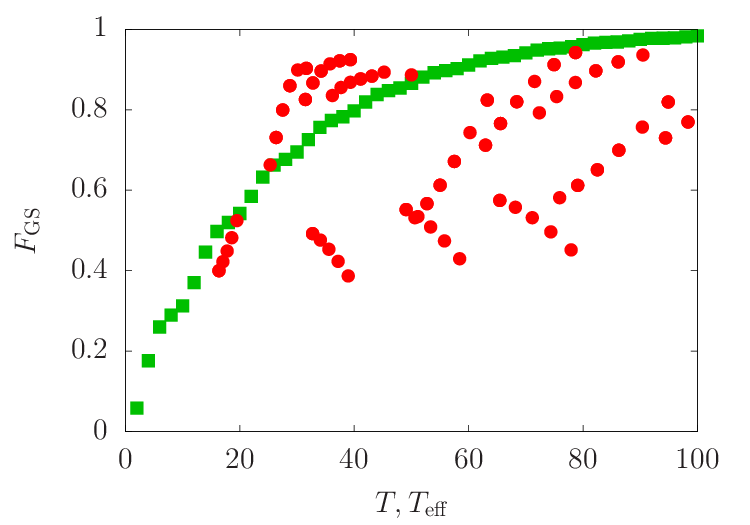}\\
\includegraphics[width=7cm]{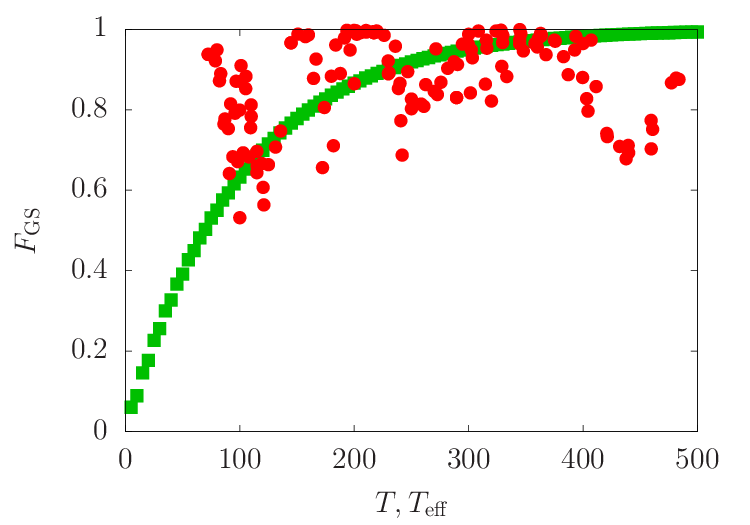}\\
\includegraphics[width=7cm]{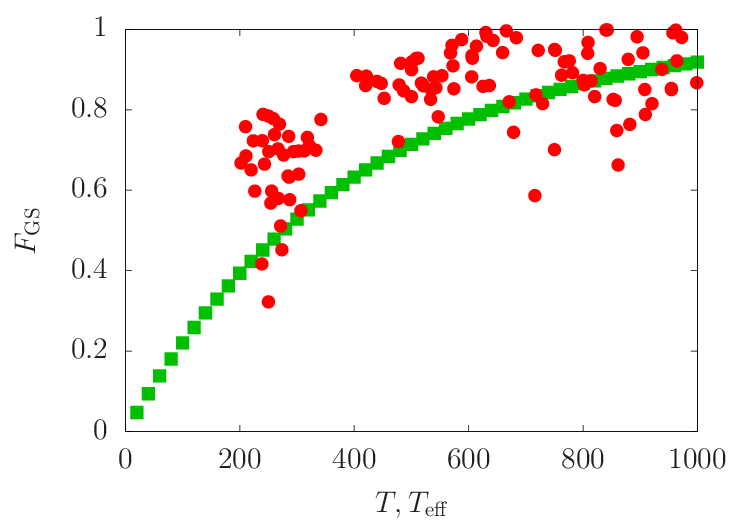}
\caption{\label{Fig.fid.two}The fidelity to the ground state. The horizontal axis is the (effective) operation time in units of $h_0^{z-1}$. The red circles represent the alternating unitary method and the green squares represent adiabatic driving. The energy gap is given by $\Delta_\mathrm{min}=0.4$ (top), $0.2$ (middle), and $0.1$ (bottom) in units of $h_0^z$. }
\end{figure}

In numerical simulation, we consider $h_0^x=0.05, 0.1, 0.2$ [$\Delta_\mathrm{min}=0.1, 0.2, 0.4$] in units of $h_0^z$. 
In the alternating unitary method, we have some parameters, $\eta$, $M$, and $L$. 
We set $\eta=(0.8+0.04\times j)\times\Delta_\mathrm{min}$ for $j=0,1,2,\dots,5$, $M=\lfloor(1+0.2\times k)/\eta\rfloor$ for $k=0,1,2,\dots,5$, and $L=2,4,6,8$. 
The fidelity to the ground state against the (effective) operation time is shown in Fig.~\ref{Fig.fid.two}. 
Here, the red circles represent the alternating unitary method and the green squares represent adiabatic driving. 
Note that some points may be located outside of the frame. 
We find that the alternating unitary method tends to outperform adiabatic driving when the energy gap is small, whereas the tendency is opposite for the large energy gap. 
We expect that this tendency also holds in other models.

%
%--------------------------------------------------------------------------
%
\subsection{$p$-spin model}

To examine the above expectation, we consider quantum annealing of the $p$-spin model described by a Hamiltonian
\begin{equation}
\hat{H}(\bm{\lambda})=-\lambda\frac{J}{N^{p-1}}\left(\sum_{i=1}^N\hat{Z}_i\right)^p-(1-\lambda)\Gamma\sum_{i=1}^N\hat{X}_i
\end{equation}
where $N$ is the number of spins, and $\hat{X}_i$ and $\hat{Z}_i$ are the Pauli-X operator and the Pauli-Z operator of the $i$th spin, respectively. 
A controllable parameter is the one-dimensional annealing parameter $\bm{\lambda}=\lambda$, and the coupling strength $J$ and the amplitude of a transverse field $\Gamma$ are uncontrollable. 
Hereafter, we set $J/\Gamma=1$. 
We realize quantum annealing by $\lambda_i=0$ and $\lambda_f=1$. 
We adopt the linear scheduling
\begin{equation}
\lambda(s)=s=t/T,
\end{equation}
for state transfer by adiabatic driving. 
The energy gap is given by $\Delta_\mathrm{min}\approx N\times N^{-1/3}$ for $p=2$~\cite{Botet1982,Caneva2008} and $\Delta_\mathrm{min}\approx N\times2^{-0.126N}$ for $p=3$~\cite{Jorg2010} in units of $J$, namely, the $p$-spin model shows the second-order transition with polynomial gap closing (per single spin, i.e., $\Delta_\mathrm{min}/N\approx N^{-1/3}$) for $p=2$ and the first-order transition with exponential gap closing for $p=3$. 
It means that the case of $p=3$ is an exponentially hard problem for quantum annealing.

In numerical simulation, we adopt the following parameters: $\eta=0.025\times(0.8+0.04\times j)\times\Delta$ with $\Delta=N\times N^{-1/3}$ for $j=0,1,2,\dots,5$, $M=N\times[(1+0.2\times k)/\eta]$ for $k=0,1,2,\ldots,5$, and $L=2,4,6,8$. 
Here, the value of $\Delta$ is chosen to match the energy gap of the $p$-spin model for $p=2$. 
We also adopt the same parameters for $p=3$ because it is not realistic to consider a parameter range for the exponentially small energy gap.

%
%+++++++++++++++++++++++++++++++++++++++++++++++++++++++++++++++++++
%
\subsubsection{$p=3$: Exponentially small energy gap}

\begin{figure}
\includegraphics[width=7cm]{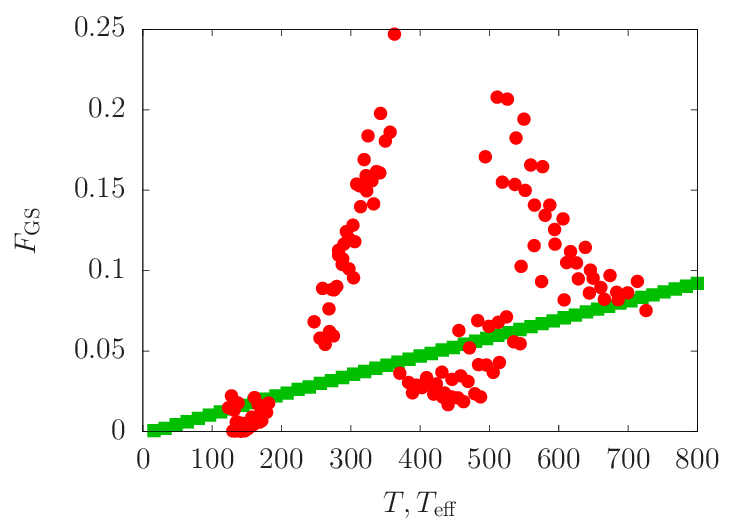}
\includegraphics[width=7cm]{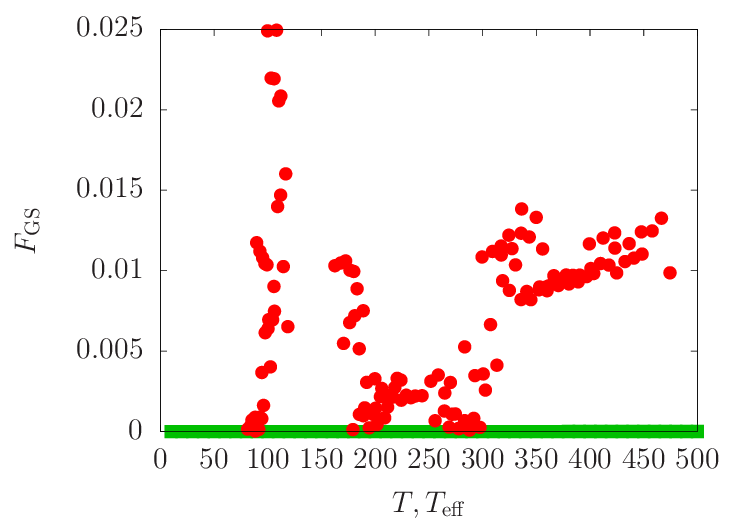}
\caption{\label{Fig.pspin}The fidelity to the ground state. The horizontal axis is the (effective) operation time in units of $J^{-1}$. The red circles represent the alternating unitary method and the green squares represent adiabatic driving. The system size is given by $N=50$ (left) and $N=100$ (right). }
\end{figure}

We discuss the case of $p=3$ with the exponentially small energy gap. 
Here, the size of the energy gap is $\Delta_\mathrm{min}\approx0.635$ for $N=50$ and $\Delta_\mathrm{min}\approx0.0161$ for $N=100$. 
The fidelity to the ground state against the (effective) operation time is shown in Fig.~\ref{Fig.pspin}. 
The red circles represent the alternating unitary method and the green squares represent adiabatic driving. 
The alternating unitary method clearly tends to outperform adiabatic driving. 
In particular, the alternating unitary method has finite fidelity to the ground state for $N=100$, while adiabatic driving only gives negligibly small fidelity.

\begin{figure}
\includegraphics[width=7cm]{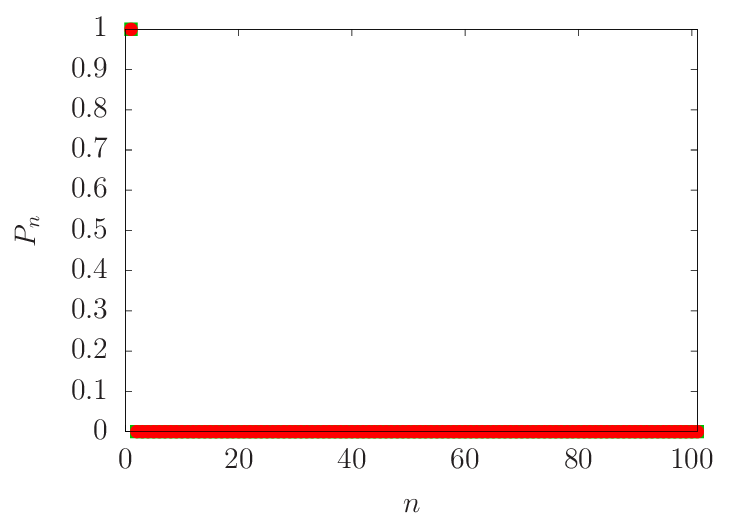}
\includegraphics[width=7cm]{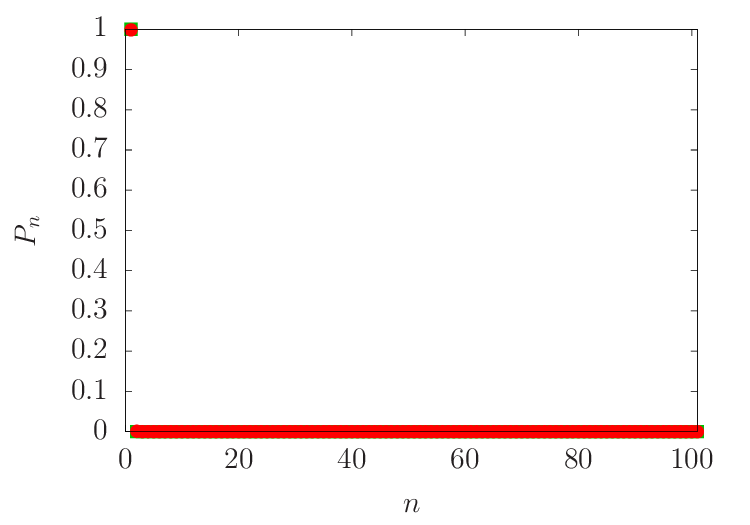}
\includegraphics[width=7cm]{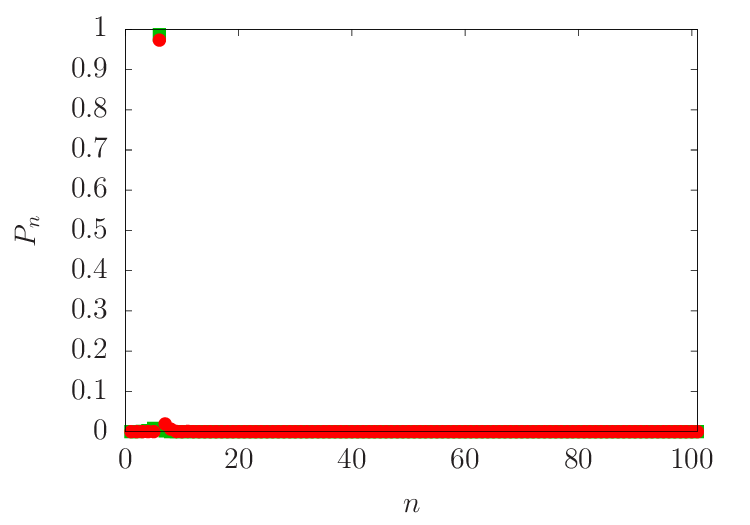}
\includegraphics[width=7cm]{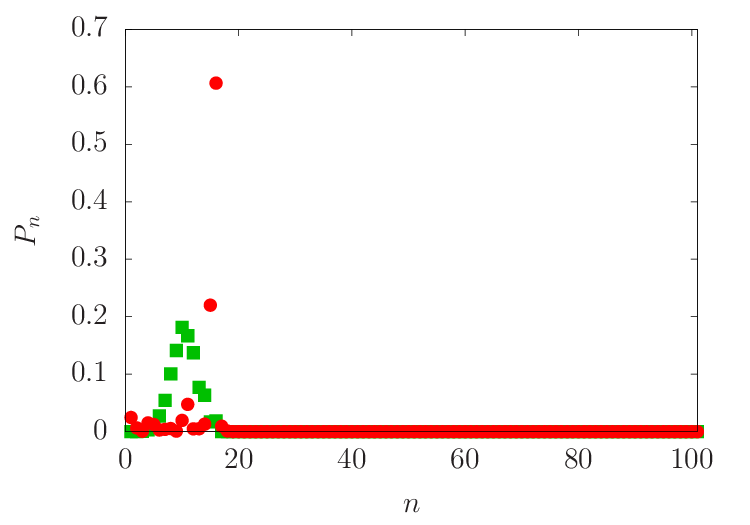}
\includegraphics[width=7cm]{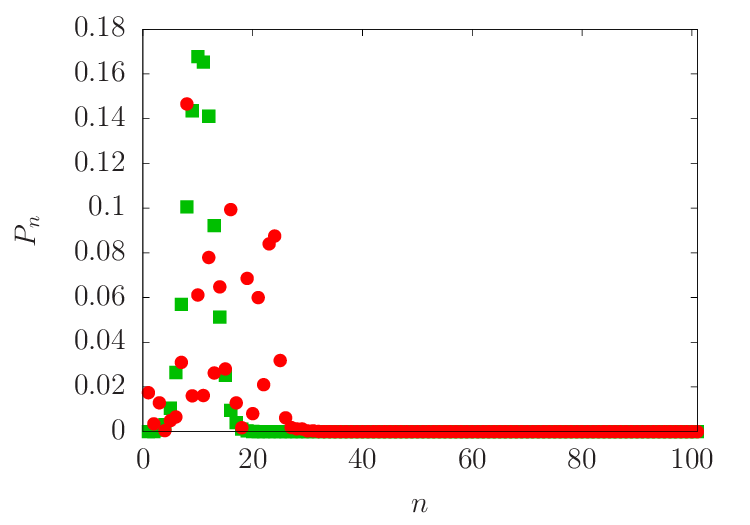}
\includegraphics[width=7cm]{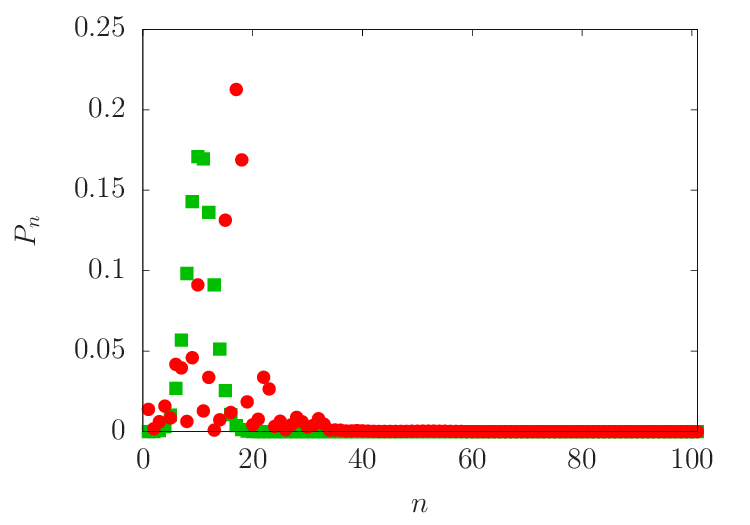}
\caption{\label{Fig.pspin.dist.6}Populations on the energy eigenstates at each parameter slice of the case No.~10 of TABLE~\ref{Tab.good}. The horizontal axis is the number of eigenstates from the ground state to the highest energy eigenstate. The red circle represent the alternating unitary method and the green squares represent adiabatic driving. Here, $\lambda=1/6$ (top left), $2/6$ (top right), $3/6$ (middle left), $4/6$ (middle right), $5/6$ (bottom left), and $6/6$ (bottom right). }
\end{figure}

From derivation, we expect that high fidelity is obtained when the regularizer $\eta$ is small (the integer $j$ is small), the number of the steps $M$ is large (the integer $k$ is large), and the number of the parameter slices $L$ is large. 
Here, we list the twenty best results of the alternating unitary method for $N=100$ in TABLE~\ref{Tab.good}. 
\begin{table}
\centering
\tabcolsep=0.5cm
\begin{tabular}{c|ccccc}
No. & $F_\mathrm{GS}$ & $T_\mathrm{eff}$ & $j$ for $\eta$ & $k$ for $M$ & $L$ \\
\hline
1 & 0.0249 & 108.14 & 1 & 3 & 2 \\
2 & 0.0249 & 99.85 & 1 & 0 & 2 \\
3 & 0.0220 & 103.07 & 1 & 1 & 2 \\
4 & 0.0219 & 105.78 & 1 & 2 & 2 \\
5 & 0.0209 & 112.12 & 1 & 5 & 2 \\
6 & 0.0205 & 110.22 & 1 & 4 & 2 \\
7 & 0.0160 & 116.64 & 0 & 4 & 2 \\
8 & 0.0147 & 111.95 & 0 & 3 & 2 \\
9 & 0.0140 & 109.11 & 0 & 1 & 2 \\
10 & 0.0138 & 336.36 & 1 & 5 & 6 \\
11 & 0.0133 & 349.93 & 0 & 4 & 6 \\
12 & 0.0133 & 466.57 & 0 & 4 & 8 \\
13 & 0.0125 & 457.89 & 0 & 3 & 8 \\
14 & 0.0124 & 447.82 & 0 & 2 & 8 \\
15 & 0.0123 & 422.99 & 0 & 0 & 8 \\
16 & 0.0123 & 335.86 & 0 & 2 & 6 \\
17 & 0.0122 & 324.42 & 1 & 3 & 6 \\
18 & 0.0121 & 343.41 & 0 & 3 & 6 \\
19 & 0.0120 & 412.26 & 1 & 1 & 8 \\
20 & 0.0117 & 89.64 & 3 & 0 & 2 \\
\end{tabular}
\caption{\label{Tab.good}The twenty highest fidelities to the ground state in the alternating unitary method applied to the $p$-spin model with $p=3$. The fidelity to the ground state $F_\mathrm{GS}$, the effective operation time $T_\mathrm{eff}$, the integer $j$ for the regularizer $\eta$, the integer $k$ for the number of steps $M$, and the number of the parameter slices $L$ are listed. }
\end{table}
There are several exceptions which do not match the above expectation, but we first study the case No.~10 of TABLE~\ref{Tab.good} which matches the above expectation. 
In Fig.~\ref{Fig.pspin.dist.6}, we depict populations on the energy eigenstates at each parameter slice ($\lambda=1/6,2/6,\dots,6/6$). 
Here, the red circles represent the alternating unitary method and the green squares represent adiabatic driving. 
At the first two steps, both methods successfully track the ground state [Fig.~\ref{Fig.pspin.dist.6} (top left, top right)]. 
In the third step, which is just after the first-order phase transition with exponential gap closing, the states are scattered over the same excited eigenstate in both methods [Fig.~\ref{Fig.pspin.dist.6} (middle left)]. 
Note that the state of the alternating unitary method has slightly broad distribution. 
After the fourth step, the state of the alternating unitary method is distributed on higher-energy eigenstates and has broader oscillating distribution than that of adiabatic driving [Fig.~\ref{Fig.pspin.dist.6} (middle right, bottom left, bottom right)]. 
This broad oscillating distribution results in finite fidelity to the ground state in the alternating unitary method.

\begin{figure}
\includegraphics[width=7cm]{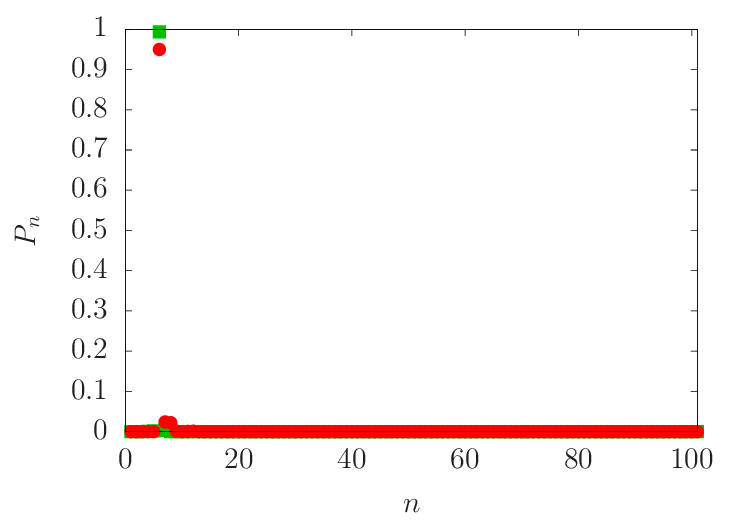}
\includegraphics[width=7cm]{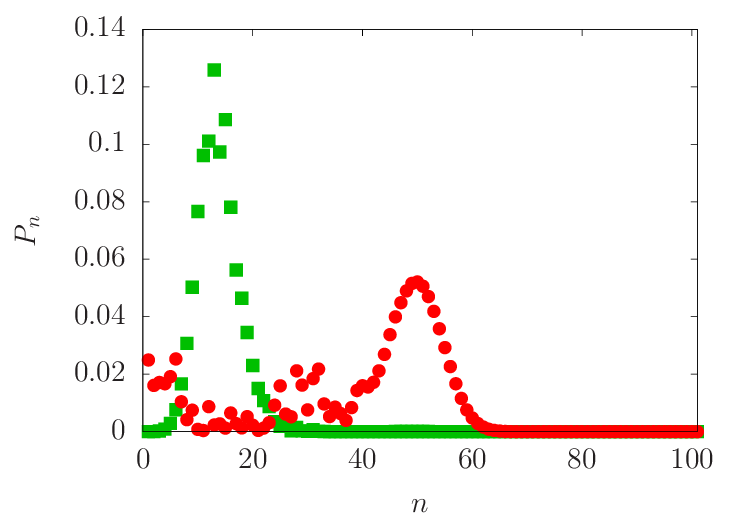}
\caption{\label{Fig.pspin.dist.2}Populations on the energy eigenstates at each parameter slice of the case No.~1 in TABLE~\ref{Tab.good}. The horizontal axis is the number of eigenstates from the ground state to the highest energy eigenstate. The red circle represent the alternating unitary method and the green squares represent adiabatic driving. Here, $\lambda=1/2$ (left) and $2/2$ (right). }
\end{figure}

We also mention exceptional cases. 
We find that exceptionally large fidelities, e.g., those in No.~1-9 and 20 of TABLE~\ref{Tab.good}, come from similar distribution to Fig.~\ref{Fig.pspin.dist.6}. 
As an example, we depict populations on the energy eigenstates at each parameter slice of the case No.~1 of TABLE~\ref{Tab.good} in Fig.~\ref{Fig.pspin.dist.2}. 
Owing to very large parameter steps (small $L$), the state of the alternating unitary method is scattered over very high-energy eigenstates, but it has a long tail in low-energy eigenstates. 
As a result, it exceptionally has high fidelity to the ground state.

%
%++++++++++++++++++++++++++++++++++++++++++++++++++++++++++++++++++++
%
\subsubsection{$p=2$: Polynomially large energy gap}

\begin{figure}
\includegraphics[width=7cm]{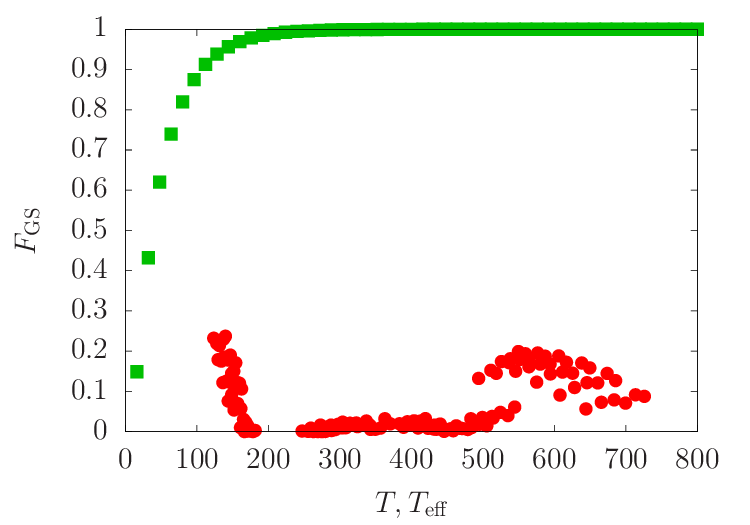}
\includegraphics[width=7cm]{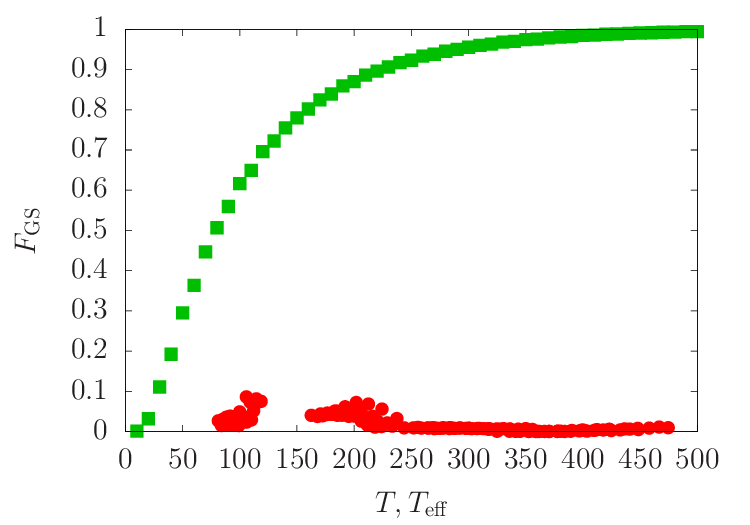}
\caption{\label{Fig.cat.fid}The fidelity to the ground state. The horizontal axis is the (effective) operation time in units of $J^{-1}$. The red circles represent the alternating unitary method and the green squares represent adiabatic driving. The system size is given by $N=50$ (left) and $N=100$ (right). }
\end{figure}

Finally, we discuss the case of $p=2$ with polynomially large energy gap. 
Here, the size of the energy gap is $\Delta_\mathrm{min}\approx13.6$ for $N=50$ and $\Delta_\mathrm{min}\approx21.5$ for $N=100$. 
The fidelity to the ground state against the (effective) operation time is plotted in Fig.~\ref{Fig.cat.fid}. 
The red circles represent the alternating unitary method and the green squares represent adiabatic driving. 
We find that adiabatic driving is much better than the alternating unitary method in this model with the polynomially large energy gap.

\begin{figure}
\includegraphics[width=7cm]{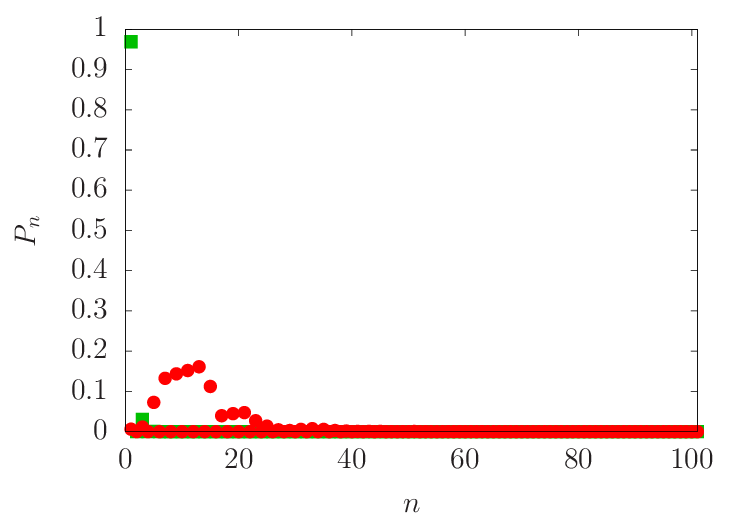}
\caption{\label{Fig.cat.dist}Populations on the energy eigenstates. The horizontal axis is the number of eigenstates from the ground state to the highest energy eigenstate. The red circles represent the alternating unitary method and the green squares represent adiabatic driving. The parameters are $j=1$ for $\mu$, $k=5$ for $M$, and $L=6$. Here, $\lambda=6/6$. }
\end{figure}

We also depict the final distribution in Fig.~\ref{Fig.cat.dist}. 
Here, the parameters are $j=1$ for $\mu$, $k=5$ for $M$, and $L=6$, which are same with the case of Fig.~\ref{Fig.pspin.dist.6}. 
We find that the distribution of the alternating unitary method is located in low-energy eigenstates, but that of adiabatic driving has very high fidelity to the ground state, and thus adiabatic driving is much better than the alternating unitary method.

%
%=======================================================================
%
\section{\label{Sec.sum}Summary and discussion}

In this paper, we conducted numerical benchmarking of the alternating unitary method, in which adiabatic transformation is approximated as the alternating unitary operators of a Hamiltonian and its parameter derivative. 
By applying it to the two-level system, we pointed out possibility that the alternating unitary method may outperform adiabatic driving when it is applied to systems with small energy gaps. 
We examined this expectation by considering quantum annealing of the $p$-spin model. 
We found that the present method outperforms adiabatic driving for $p=3$, which has the exponentially small energy gap, while adiabatic driving outperforms the present method for $p=2$, which has the polynomially large energy gap. 
This result is a consequence of broad distribution on energy eigenstates. 
The present result indicates that the alternating unitary method may have the ability to sample low-energy eigenstates in quantum annealing applied to hard instances.

It should be stressed that the alternating unitary method does not require additional interaction terms in contrast to counterdiabatic driving, whereas the derivation of the present method is clearly related to the additional driving term of counterdiabatic driving. 
Accordingly, the present method would be experimentally realizable by using programmable quantum devices, e.g., quantum annealers or quantum processors.

Comparison between the present result and counterdiabatic driving with practical approximation would be of great interest. 
In Ref.~\cite{Prielinger2021}, application of local counterdiabatic driving based on the variational method~\cite{Sels2017} to quantum annealing of $p$-spin model with $p=3$ was discussed. 
They also introduced a two-parameter approach, which shows better performance than the simple application of the variational method. 
As shown in Fig.~1 of Ref.~\cite{Prielinger2021}, for any method they studied, the fidelity to the ground state is smaller than $0.1$ for the annealing time $T=1000$ and the system size $N=50$. 
In contrast, we can obtain the fidelity about $0.25$ for the effective operation time $T_\mathrm{eff}\approx363$ in the present method as shown in Fig.~\ref{Fig.pspin}. 
Note that we can also find application of counterdiabatic driving with the Holstein-Primakoff approximation to a model which is equivalent to quantum annealing of the $p$-spin model with $p=2$ in Ref.~\cite{Hatomura2018a}. 
It was shown that approximate counterdiabatic driving is better than adiabatic driving. 
Therefore, the alternating unitary method may be the best choice for systems with exponentially small energy gap among the alternating unitary method, approximate counterdiabatic driving, and adiabatic driving, while approximate counterdiabatic driving might be the best choice for systems with polynomially large energy gaps among these methods.

\begin{acknowledgments}
This work was supported by JST Moonshot R\&D Grant Number JPMJMS2061.
The author is grateful to Dyon van Vreumingen for fruitful discussion. 
\end{acknowledgments}

\appendix
\section*{Appendix}

%
%======================================================================
%
\section{\label{Sec.Appendix.reg.AGP}Regularized adiabatic gauge potential}

The $(m,n)$ element of the adiabatic gauge potential (\ref{Eq.AGP}) is given by
\begin{equation}
\begin{aligned}
\langle m(\bm{\lambda})|\hat{\mathcal{A}}(\bm{\lambda})|n(\bm{\lambda})\rangle&=i\hbar\langle m(\bm{\lambda})|\bm{\nabla}n(\bm{\lambda})\rangle\\
&=i\hbar\frac{\langle m(\bm{\lambda})|\bm{(}\bm{\nabla}\hat{H}(\bm{\lambda})\bm{)}|n(\bm{\lambda})\rangle}{E_n(\bm{\lambda})-E_m(\bm{\lambda})}. 
\end{aligned}
\end{equation}
and that of the regularized adiabatic gauge potential (\ref{Eq.reg.AGP}) is given by
\begin{equation}
\langle m(\bm{\lambda})|\hat{\mathcal{A}}_\mu(\bm{\lambda})|n(\bm{\lambda})\rangle=i\hbar\frac{E_n(\bm{\lambda})-E_m(\bm{\lambda})}{\hbar^2\eta^2+\bm{(}E_n(\bm{\lambda})-E_m(\bm{\lambda})\bm{)}^2}\langle m(\bm{\lambda})|\bm{(}\bm{\nabla}\hat{H}(\bm{\lambda})\bm{)}|n(\bm{\lambda})\rangle. 
\end{equation}
Therefore, it is a good approximation when $\hbar^2\eta^2\ll\bm{(}E_n(\bm{\lambda})-E_m(\bm{\lambda})\bm{)}^2$ is satisfied, and $\lim_{\eta\to0}\hat{\mathcal{A}}_\eta(\bm{\lambda})=\hat{\mathcal{A}}(\bm{\lambda})$ holds.

%
%======================================================================
%
\section{\label{Sec.disc.reg.AGP}Discretization of the regularized adiabatic gauge potential}

The regularized adiabatic gauge potential (\ref{Eq.reg.AGP}) can be discretized as
\begin{equation}
\begin{aligned}
\hat{\mathcal{A}}_\eta(\bm{\lambda})=&-\frac{1}{2}\int_0^\infty ds\ e^{-\eta s}\left(e^{\frac{i}{\hbar}\hat{H}(\bm{\lambda})s}\bm{(}\bm{\nabla}\hat{H}(\bm{\lambda})\bm{)}e^{-\frac{i}{\hbar}\hat{H}(\bm{\lambda})s}-e^{-\frac{i}{\hbar}\hat{H}(\bm{\lambda})s}\bm{(}\bm{\nabla}\hat{H}(\bm{\lambda})\bm{)}e^{\frac{i}{\hbar}\hat{H}(\bm{\lambda})s}\right) \\
=&\frac{1}{2\eta}\int_1^0du\left(u^{-\frac{i}{\hbar\eta}\hat{H}(\bm{\lambda})}\bm{(}\bm{\nabla}\hat{H}(\bm{\lambda})\bm{)}u^{\frac{i}{\hbar\eta}\hat{H}(\bm{\lambda})}-u^{\frac{i}{\hbar\eta}\hat{H}(\bm{\lambda})}\bm{(}\bm{\nabla}\hat{H}(\bm{\lambda})\bm{)}u^{-\frac{i}{\hbar\eta}\hat{H}(\bm{\lambda})}\right) \\
\approx&\frac{1}{2\eta M}\sum_{m=M-1}^{1}\left[\left(\frac{m}{M}\right)^{-\frac{i}{\hbar\eta}\hat{H}(\bm{\lambda})}\bm{(}\bm{\nabla}\hat{H}(\bm{\lambda})\bm{)}\left(\frac{m}{M}\right)^{\frac{i}{\hbar\eta}\hat{H}(\bm{\lambda})}-\left(\frac{m}{M}\right)^{\frac{i}{\hbar\eta}\hat{H}(\bm{\lambda})}\bm{(}\bm{\nabla}\hat{H}(\bm{\lambda})\bm{)}\left(\frac{m}{M}\right)^{-\frac{i}{\hbar\eta}\hat{H}(\bm{\lambda})}\right] \\
=&\frac{1}{2\eta M}\sum_{m=M-1}^{1}\left(e^{-\frac{i}{\hbar\eta}\log(m/M)\hat{H}(\bm{\lambda})}\bm{(}\bm{\nabla}\hat{H}(\bm{\lambda})\bm{)}e^{\frac{i}{\hbar\eta}\log(m/M)\hat{H}(\bm{\lambda})}-e^{\frac{i}{\hbar\eta}\log(m/M)\hat{H}(\bm{\lambda})}\bm{(}\bm{\nabla}\hat{H}(\bm{\lambda})\bm{)}e^{-\frac{i}{\hbar\eta}\log(m/M)\hat{H}(\bm{\lambda})}\right) \\
=&\frac{1}{2\eta M}\sum_{\substack{m=-M+1 \\ (m\neq0)}}^{M-1}\mathrm{sgn}(m)e^{-\frac{i}{\hbar\eta}\mathrm{sgn}(m)\log|m/M|\hat{H}(\bm{\lambda})}\bm{(}\bm{\nabla}\hat{H}(\bm{\lambda})\bm{)}e^{\frac{i}{\hbar\eta}\mathrm{sgn}(m)\log|m/M|\hat{H}(\bm{\lambda})} \\
\equiv&\hat{\mathcal{A}}_\eta^M(\bm{\lambda}),
\end{aligned}
\label{Eq.appendix.appAGP}
\end{equation}
where we set $u=e^{-\eta s}$ in the second line and adopt the trapezoidal rule of integral in the third line. 
Note that a cutoff was introduce in Ref.~\cite{vanVreumingen2024} to avoid integral over the infinite range, but we avoid it by introducing the variable conversion $u=e^{-\eta s}$. 
In addition, we adopt the trapezoidal rule of integral for simplicity, whereas weighted sum was adopted in Ref.~\cite{vanVreumingen2024}.

Then, we consider parallel transport by the above approximate adiabatic gauge potential (\ref{Eq.appendix.appAGP}). 
We obtain
\begin{equation}
\begin{aligned}
&\exp\left(-\frac{i}{\hbar}\delta\bm{\lambda}\cdot\hat{\mathcal{A}}_\eta^M(\bm{\lambda})\right)\\
=&\exp\left[-\frac{i}{2\hbar\eta M}\left(\sum_{m=-1}^{-M+1}+\sum_{m=M-1}^1\right)\mathrm{sgn}(m)e^{-\frac{i}{\hbar\eta}\mathrm{sgn}(m)\log|m/M|\hat{H}(\bm{\lambda})}\bm{(}\delta\bm{\lambda}\cdot\bm{\nabla}\hat{H}(\bm{\lambda})\bm{)}e^{\frac{i}{\hbar\eta}\mathrm{sgn}(m)\log|m/M|\hat{H}(\bm{\lambda})}\right] \\
\approx&\left(\prod_{m=-1}^{-M+1}\prod_{m=M-1}^1\right)\exp\left(\frac{i}{2\hbar\eta M}\mathrm{sgn}(m)e^{-\frac{i}{\hbar\eta}\mathrm{sgn}(m)\log|m/M|\hat{H}(\bm{\lambda})}\bm{(}\delta\bm{\lambda}\cdot\bm{\nabla}\hat{H}(\bm{\lambda})\bm{)}e^{\frac{i}{\hbar\eta}\mathrm{sgn}(m)\log|m/M|\hat{H}(\bm{\lambda})}\right) \\
=&\left(\prod_{m=-1}^{-M+1}\prod_{m=M-1}^1\right)e^{-\frac{i}{\hbar\eta}\mathrm{sgn}(m)\log|m/M|\hat{H}(\bm{\lambda})}e^{\frac{i}{2\hbar\eta M}\mathrm{sgn}(m)\bm{(}\delta\bm{\lambda}\cdot\bm{\nabla}\hat{H}(\bm{\lambda})\bm{)}}e^{\frac{i}{\hbar\eta}\mathrm{sgn}(m)\log|m/M|\hat{H}(\bm{\lambda})} \\
=&e^{\frac{i}{\hbar\eta}\log(1/M)\hat{H}(\bm{\lambda})}\left(\prod_{m=-1}^{-M+1}e^{-\frac{i}{2\hbar\eta M}\bm{(}\delta\bm{\lambda}\cdot\bm{\nabla}\hat{H}(\bm{\lambda})\bm{)}}e^{-\frac{i}{\hbar\eta}\log\bm{(}m/(m-1)\bm{)}\hat{H}(\bm{\lambda})}\right)\\
&\times\left(\prod_{m=M-1}^1e^{-\frac{i}{\hbar\eta}\log\bm{(}m/(m+1)\bm{)}\hat{H}(\bm{\lambda})}e^{\frac{i}{2\hbar\eta M}\bm{(}\delta\bm{\lambda}\cdot\bm{\nabla}\hat{H}(\bm{\lambda})\bm{)}}\right)e^{\frac{i}{\hbar\eta}\log(1/M)\hat{H}(\bm{\lambda})}\\
\equiv&\hat{u}_\eta^M(\bm{\lambda})\hat{U}_\eta^M(\bm{\lambda},\delta\bm{\lambda})\hat{u}_\eta^M(\bm{\lambda})\\
\equiv&\hat{\mathcal{U}}_\eta^M(\bm{\lambda},\delta\bm{\lambda}),
\end{aligned}
\label{Eq.appendix.apptrans}
\end{equation}
where we use Trotterization in the third line.

%
%------------------------------------------------------------------------
%
\section{\label{Sec.reduc.time}Reduction of the effective operation time}

We consider reduction of the effective operation time. 
By changing the order of summation
\begin{equation}
\sum_{m=-1}^{-M+1}+\sum_{m=M-1}^1\to\sum_{m=1}^{M-1}+\sum_{m=-M+1}^{-1},
\end{equation}
and the order of product
\begin{equation}
\prod_{m=-1}^{-M+1}\prod_{m=M-1}^1\to\prod_{m=1}^{M-1}\prod_{m=-M+1}^{-1},
\end{equation}
in the above calculation (\ref{Eq.appendix.apptrans}), we also obtain
\begin{equation}
\begin{aligned}
\exp\left(-\frac{i}{\hbar}\delta\bm{\lambda}\cdot\hat{\mathcal{A}}_\eta^M(\bm{\lambda})\right)\approx&\hat{u}_\eta^{M\dag}(\bm{\lambda})\hat{U}_\eta^{M\dag}(\bm{\lambda},-\delta\bm{\lambda})\hat{u}_\eta^{M\dag}(\bm{\lambda})\\
=&\hat{\mathcal{U}}_\eta^{M\dag}(\bm{\lambda},-\delta\bm{\lambda}).  
\end{aligned}
\end{equation}
Then, we can consider the following unitaries
\begin{equation}
\prod_{n=L/2-1}^{0}\hat{\mathcal{U}}_\eta^{M\dag}\bm{(}\bm{\lambda}_i+(2n+1)\delta\bm{\lambda},-\delta\bm{\lambda}\bm{)}\hat{\mathcal{U}}_\eta^{M}(\bm{\lambda}_i+2n\delta\bm{\lambda},\delta\bm{\lambda})
\end{equation}
for state transfer from $|n(\bm{\lambda}_i)\rangle$ to $|n(\bm{\lambda}_f)\rangle$. 
It may be advantageous because we can expect
\begin{equation}
\hat{u}_\eta^M(\bm{\lambda}+\delta\bm{\lambda})\hat{u}_\eta^{M\dag}(\bm{\lambda})\approx e^{\frac{i}{\hbar\eta}\log(1/M)\bm{(}\delta\bm{\lambda}\cdot\bm{\nabla}\hat{H}(\bm{\lambda})\bm{)}}
\end{equation}
for a small parameter step $|\delta\bm{\lambda}|\ll1$. 
Moreover, we could neglect the first and final factors $\hat{u}_\eta^{M}(\bm{\lambda}_i)$ and $\hat{u}_\eta^{M\dag}\bm{(}\bm{\lambda}_i+(L-1)\delta\bm{\lambda}\bm{)}$ because these two unitaries are expected as global phase factors for $|n(\bm{\lambda}_i)\rangle$ and $|n(\bm{\lambda}_f)\rangle$. 
Finally, we obtain a reduced effective operation time
\begin{equation}
T_\mathrm{eff}^{r}=\frac{3L-1}{\eta}\log M+\frac{L}{\eta}\left(1-\frac{1}{M}\right). 
\end{equation}
That is, we can reduce the effective operation time by $[(L+1)/\eta]\log M$.

\bibliography{altunitarybib}

\end{document}